%% file: main.tex
\documentclass[conference, 11pt]{IEEEtran}
\IEEEoverridecommandlockouts

\usepackage{graphicx,color,placeins,float,tabularx,colortbl,amsmath,amssymb,epsfig,amsthm,cite,dblfloatfix}
\usepackage[pdfusetitle, pdfauthor={Tarek Aziz Lahlou, MIT}]{hyperref}
\hypersetup{colorlinks = true,allcolors=black}

\input{definitions.tex}
\begin{document} \onecolumn

\title{Web Services for Asynchronous, Distributed Optimization Using Conservative Signal Processing}

\author{\IEEEauthorblockN{Tarek A. Lahlou and Thomas A. Baran}
\IEEEauthorblockA{Digital Signal Processing Group  \\ Massachusetts Institute of Technology }}

\maketitle

\begin{abstract}
	This paper presents a systematic approach for implementing a class of nonlinear signal processing systems as a distributed web service, which in turn is used to solve optimization problems in a distributed, asynchronous fashion.   As opposed to requiring a specialized server, the presented approach requires only the use of a commodity database back-end as a central resource, as might typically be used to serve data for websites having large numbers of concurrent users.  In this sense the presented approach leverages not only the scalability and robustness of various database systems in sharing variables asynchronously between workers, but also critically it leverages the tools of signal processing in determining how the optimization algorithm might be organized and distributed among various heterogeneous workers.  A publicly-accessible implementation is also  presented, utilizing Firebase as a back-end server, and illustrating the use of the presented approach in solving various optimization problems commonly arising in the context of signal processing.
\end{abstract}

\tableofcontents

\newpage 

\section{Introduction} \label{sec:intro}
	In designing and implementing signal processing systems, a general implementation strategy is to (1) begin with a set of desired equations to be satisfied, (2) represent these equations as a graphical structure, (3) distribute state throughout the graph, e.g.~introducing scalar or vector delay elements, and (4) determine a protocol for exchanging state, resulting in an algorithm or iteration satisfying the original equations.  Various specific methods consistent with this general approach have been described formally, e.g.~in \cite{BaranLahlouICASSP15, Crochiere}.

	Consistent with these steps, the issue of distributing state is perhaps the most central issue in effectively distributing algorithms in general, including distributing algorithms across multiple heterogeneous processing nodes.  As has been discussed in \cite{NikkilArvind}, this observation suggests opportunity in utilizing the general strategy of specifying algorithms first using a declarative language, which after determining a protocol for distributing and exchanging state, would be decomposed as an ensemble of distributed programs and implemented on processing nodes using imperative languages.

	The formal approaches used in implementing signal processing systems form a broad and concrete class of examples that are consistent with this general strategy, with state-free signal-flow diagrams being a declarative representation, and with an eventual arrangement of run-loops being imperative.  Drawing upon this, the results outlined in \cite{tbaran-phd, BLOpt, LahlouBaranLinProg} describe a straightforward method for implementing a variety of \emph{optimization algorithms} by casting them as signal processing systems, in turn leveraging the various common associated implementation strategies in distributing and transferring~state.

	The intent of this paper is to describe a distributed web service for solving optimization problems that results as a consequence of the way of thinking described in \cite{BLOpt, LahlouBaranLinProg, tbaran-phd}.  The service is freely accessible online as part of the general site ``Signal Processing Conservation''\cite{spconservation}, which provides a general overview and examples of the use of conservation principles in signal processing systems, importantly also describing the mathematical foundation underlying \cite{BLOpt, LahlouBaranLinProg, tbaran-phd}.  The portion of the site containing the web service for optimization discussed in this paper is available at http://optimization.spconservation.org, which we refer to herein as ``O-SPC''.

	The architecture of O-SPC in particular is built on Firebase \cite{firebase} and utilizes the service primarily as a high-performance back-end for asynchronous representational state transfer between browser-based clients, e.g.~as opposed to as a centralized resource for coordinating data processing as with \cite{parameterServer}.  In this sense, O-SPC represents an example of how the thinking described in \cite{BLOpt, LahlouBaranLinProg, tbaran-phd} can be used to create a performant system operating in the somewhat extreme case where numerical computation is distributed entirely to the extremities of the graph.  The considerations described in this paper would similarly apply to the creation of a web-based optimization service utilizing an alternative key-value store system, e.g.~MongoDB \cite{MongoDB} or Redis \cite{Redis}, or any number of relational database systems.  In each of these cases, the performance of the distributed system would be able to draw upon the particular strengths of the data store being utilized.
	
	We begin in Section~\ref{sec:sig-flow} by specifying the targeted class of signal processing systems and reviewing their utility as optimization algorithms. In Section~\ref{sec:opt.spc} we focus on general considerations regarding their distributed implementation as a web service, consistent with the architecture of O-SPC. In Section~\ref{sec:examples}, we collate the numerical experiments referenced throughout and provide concluding remarks.

	\begin{figure*}[b!]
		\setcounter{figure}{0}
		\centering
		\centerline{\includegraphics[width=\textwidth]{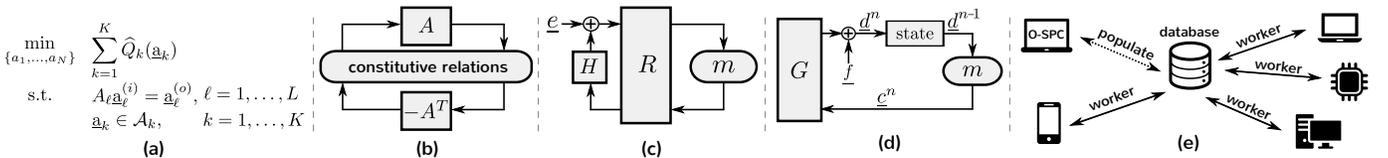}}
		\caption{An illustration of (a) the primal optimization problem, (b) the associated stationarity conditions derived from \cite{tbaran-phd}, and (c) the transformed stationarity conditions in \cite{BLOpt}. (d) The implementation of \eqref{eq:c-and-d} utilized in \cite{BLOpt2} obtained by distributing state to form an algorithm. (e) The compute architecture leveraged by O-SPC by distributing state via a database: a client populates the database with a problem instance and then uncoordinated workers asynchronously implement the conditions in \eqref{eq:c-and-d}.}
		\label{fig:signal-flow}
	\end{figure*}

\section{Signal processing systems and optimization}\label{sec:sig-flow} 
	The general framework presented in \cite{BLOpt} facilitates the construction of distributed, asynchronous signal processing systems for solving optimization problems by analyzing the structure of the optimization problem itself and without relying on any existing non-distributed and/or synchronous methods. Therefore using this framework, signal processing systems, and by extension optimization algorithms, may be generated that might not be readily derived by conventional techniques. In the remainder of this section we briefly review the key steps in casting optimization algorithms as signal processing systems.

	A conservative signal processing system is one for which the variables available for interconnection between subsystems admit an organization adhering to an indefinite quadratic form of a particular class that is invariant to the evolution of the system \cite{tbaran-phd}. The utility of conservation principles in \cite{BLOpt} is twofold: (1) in defining the primal optimization problem in Fig.~\ref{fig:signal-flow}(a) and its dual so that the joint feasibility conditions depicted in Fig.~\ref{fig:signal-flow}(b) serve as sufficient conditions for stationarity, and (2) in transforming said conditions into the algebraic form illustrated in Fig.~\ref{fig:signal-flow}(c) where $R:\Rn \to \Rn$ and $H\colon \R^{N-K}\to\R^{N-K}$ are orthogonal matrices, $m \colon \Rk \to \Rk$ is a generally nonlinear map, and $\e\in\R^{N-K}$ is a system bias. The maps $m$ and $H$ as well as the bias $\e$ are associated with the set constraints $\mathcal{A}_{k}$ and objective functionals $\widehat{Q}_{k}$ in Fig.~\ref{fig:signal-flow}(a) defined on the decision variables $a_k$ while $R$ is given by
	\begin{equation}
		R = \begin{bmatrix} I & -A^{T} \\ A & I \end{bmatrix}^{2}\begin{bmatrix}\left(I + A^{T}A \right)^{-1} & 0 \\ 0 & \left(I + AA^{T} \right)^{-1} \end{bmatrix} \label{eq:Q}
	\end{equation}
	where $A$ represents the aggregate linear equality constraints $A_\ell$ involving only the primal decision variables. Without loss of generality, the system in Fig.~\ref{fig:signal-flow}(c) may be recast into an equivalent system, in the sense that a solution to one yields a solution to both, of the form
	\begin{eqnarray}
		\c^\star  =  m(\d^\star)  & \text{ and } & \d^\star  =  G\c^\star + \f \label{eq:c-and-d} 
	\end{eqnarray}
	where $\c^\star,\d^\star \in \Rk$ denote a solution, $G:\Rk \to \Rk$ is an orthogonal matrix, and $\f \in \Rk$ is a system bias. Figure~\ref{fig:signal-flow}(d) illustrates the reduced system \eqref{eq:c-and-d} utilized in \cite{BLOpt2} where the algebraic loops have been broken by inserting state/memory into the system. 

	The precompute required to assemble a signal processing system of the presented class is analytic and involves purely linear operations. In particular, aside from the computation of $R$ in \eqref{eq:Q}, the algebraic reduction of $(R,H,\e)$ to $(G,\f)$ corresponds to identifying the intersection of affine subspaces and thus can be expressed in closed form. The postcompute associated with recovering the solution to the optimization problem given a solution $(\c^\star,\d^\star)$  to \eqref{eq:c-and-d} is also linear. For example, let $\a_j$ denote a primal decision variable and assume the precompute retains the system variables $\c_j$ and $\d_j$ associated with $\a_j$. Then, the value $\a_j^\star$ at a stationary point $\a^\star$ of the problem is
	\begin{eqnarray} \label{eq:readout}
		\a^{\star}_{j} = \frac{1}{2}\left(\d^\star_j + \c^\star_j\right) & \text{ or } & \a^{\star}_{j} = \frac{1}{2}\left(\d^\star_j - \c^\star_j\right)
	\end{eqnarray}
	depending on whether $\a_j$ is an input to or output from $A$, respectively.

	In the context of numerically solving \eqref{eq:c-and-d} by generating state sequences $\cn$ and $\dn$, we refer to an \emph{iterative solver} as a system implementation in which the processing directly yields the next state values and an \emph{incremental solver} as one in which the processing yields values to be added to the current state in order to produce the next. We refer to either solver as being \emph{filtered} when additional processing is used to produce the next state value as an affine combination of the current state value and the state value produced by the unfiltered solver. A sufficient condition under which the state sequences converge to a solution $(\c^\star,\d^\star)$ of \eqref{eq:c-and-d} that encompasses the numerical examples presented in this paper (provided that we appropriately implement the filtered solvers) is that the nonlinear map $m$ be non-expansive, i.e.~$m$ must satisfy
	\begin{eqnarray} \label{eq:passive-everywhere}
		\forall\,\u,\v\in\Rk, & \left\|m(\v) - m(\u)\right\|_2 \leq \left\|\v - \u\right\|_2. &
	\end{eqnarray}
	Convergence is in particular in the Euclidean sense for synchronous implementations and in mean square for stochastic/asynchronous implementations; we refer to \cite{convergence} for a complete treatment. For the purpose of illustration and not by limitation, we assume hereon that $m$ is a coordinatewise nonlinearity, this assumption is true for all numerical examples in this paper. The handling of general nonlinearities follows in an analogous way.

\section{Implementation as a web service} \label{sec:opt.spc}
	We now overview the operating principle behind O-SPC which we believe suggests opportunity in designing future systems in this way. Referring to the website content, several optimization problems frequently occurring in signal processing contexts have been assembled into an examples library, including those discussed in Section~\ref{sec:examples}. 

	In the remainder of this section we present the details associated with two distributed and four non-distributed solvers used to obtain a solution $(\c^\star,\d^\star)$ to \eqref{eq:c-and-d}. We comment upfront, however, that the specific form of the solvers presented in this section differ from the implementations in O-SPC in that they have been adapted here for the purpose of clarity rather than computational efficiency. 
	
		\begin{figure*}[t]
			\setcounter{figure}{1}
			\centering
		  	\centerline{\includegraphics[width=\textwidth]{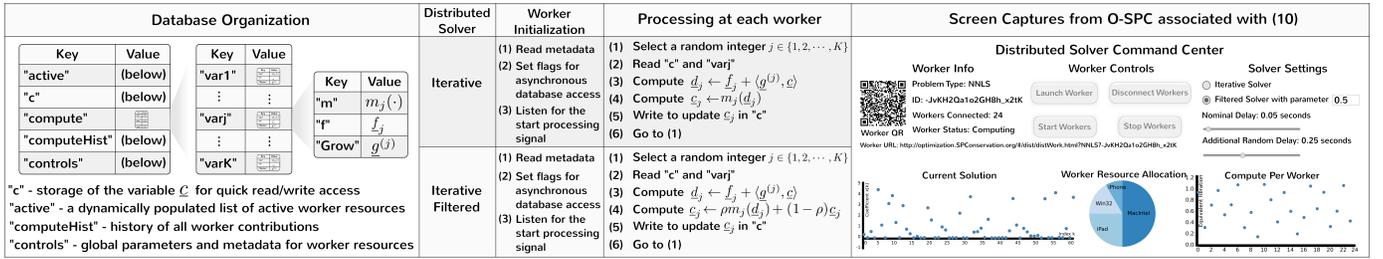}} 
		  	\caption{A qualitative description of the organization of the set of equations in \eqref{eq:c-and-d} and the signals to be processed into a generic database. The procedures for worker initialization and processing associated with two distributed solvers are also provided. Screen captures from O-SPC illustrate the global controller and computed solution of a non-negative least squares problem obtained using $24$ distributed workers implementing an iterative filtered solver with parameter $\rho = 0.5$. A breakdown of the workers computational platform allocation and individual contributions to the overall solution is also depicted.} 
			\label{fig:distributed-solvers}
		\end{figure*}

	\subsection{Distributed implementations} \label{subsec:distributed}
		A longstanding approach to efficiently solving a large class of numerical problems is to recast any problem of the class into a fixed representation to which a set of generic tools may be immediately applied. In this spirit, a signal processing system conforming to \eqref{eq:c-and-d} is automatically synthesized once the parameters of a problem have been specified, from which several solvers corresponding to various distributions of state and processing instructions may be applied.

		Consistent with the general implementation strategy discussed in Section~\ref{sec:intro}, Fig.~\ref{fig:distributed-solvers} illustrates the organization of the set of equations in \eqref{eq:c-and-d} into a generic key-value store, e.g.~a non-relational database, for implementation on the graph depicted in Fig.~\ref{fig:signal-flow}(e). The protocol for state transfer consists of workers asynchronously accessing the database to retrieve a subset of the computation and the associated signals to be processed, processing these signals, and asynchronously writing the result back into the database. This strategy represents a form of object-oriented signal processing wherein the objects contain data in the form of the signals to be processed and methods in the form of processing~instructions.

		Screen captures from the O-SPC application interface are provided in Fig.~\ref{fig:distributed-solvers} for a non-negative least squares problem, depicting the dashboard through which distributed workers can be controlled. Through the dashboard interface, metaparameters for the problem can be set, in turn generating a corresponding uniform resource locator (URL) through which workers can attach to the problem instance to perform computation. For worker devices with integrated cameras, a quick response (QR) code is also dynamically generated.	The particular solution depicted in Fig.~\ref{fig:distributed-solvers} was obtained using $24$ distributed workers.  Analytics regarding the computational platforms of the connected workers, as well as the individual contributions to the overall optimization progress, are provided via dynamically-generated graphs.

		It is worth noting that nearly any computational resource equipped with network access and a basic JavaScript engine may be utilized as a worker on O-SPC. For example, a heterogeneous set of workers might include modern web browsers on mobile, tablet and desktop machines as well as JavaScript enabled microcontrollers \cite{tessel,Espruino}.

		The worker initialization and processing instructions for two distributed solvers are summarized in column 3 of Fig.~\ref{fig:distributed-solvers}. Specifically, each worker, independent of any and all other workers, performs the following steps \emph{ad infinitum} to implement an iterative filtered solver: 
		\begin{enumerate}
			\item[(1)] generate a random integer $j\in\{1,\dots,K\}$ corresponding to the state variables $\c_j$ and $\d_j$ to be processed;
			\item[(2)] read the current state of the vector $\c$ as well as the object {\tt varj} consisting of a characterization of the nonlinearity $m_j$ labeled {\tt m}, the value of $\f_j$ labeled {\tt f}, and the row vector $\g^{(j)}$ corresponding to the $j^{\text{th}}$ row of $G$ labeled {\tt Grow}; 
			\item[(3)] generate the intermediary state value $\d_j$ as
				\begin{eqnarray} \label{eq:dist-filt-pre}
					\d_{j} \leftarrow  g^{(j)}_{1}\c_{1} + \cdots + g^{(j)}_{K}\c_{K} + \f_{j} 
				\end{eqnarray}
			\item[(4)] generate the new state value $\c_j$ as
				\begin{eqnarray} \label{eq:dist-filt}
					\c_{j} \leftarrow \rho m_{j}\left( \d_{j} \right) + (1-\rho)\c_{j} 
				\end{eqnarray}
				where the filtering parameter $\rho$ is a metaparameter obtained during the worker initialization phase; 
			\item[(5)] asynchronously write the new state value $\c_j$ into the $j^{\text{th}}$ position of $\c$ in the database.
		\end{enumerate}
		For iterative solvers, the state variable $\d$ does not need to be explicitly stored in the database. Indeed, once the partial solution $\c^\star$ is identified, the state vector $\d^\star$ may be generated using \eqref{eq:c-and-d} and thus the original optimization problem is effectively solved. Referring again to Fig.~\ref{fig:distributed-solvers}, the processing procedure for an iterative solver corresponds to modifying the instructions outlined above by setting  $\rho = 1$ in \eqref{eq:dist-filt}.
		
		We call special attention to the fact that no attempt is made at the algorithm level to regulate global task allocation among the workers nor to enforce concurrency of any form. Specifically, the data requests and updates are respectively executed using asynchronous read and write operations with no concept of precedent or preference among the workers. For example, if multiple workers request data associated with the same state variable $\c_j$ and each experiences a different latency (and thus each possibly retrieves different state vectors $\c$) then the database records the updates in the order they are received irrespective of the order of the read operations. 

		Referring to Fig.~\ref{fig:signal-flow}(e), the database might simultaneously contain numerous active problem instances. Workers may be added or removed at any time (including changing problem instances) without any form of coordination since workers are never assigned responsibility for any particular part of the workload. In this sense, O-SPC facilitates the time-varying allocation of compute resources in order to adaptively respond to real-time constraints, time-varying network congestion, and resource outages. Another advantage to utilizing the presented approach for solving optimization problems in practice is the ability to update the portion of the database (and by extension the signal processing structure as well) associated with measurements and/or observations as new data becomes available. The response of the system is then to transform the state of the database associated with the current solution toward the new fixed-point or invariant state corresponding to the new solution. Consequently, the distributed solvers summarized in Fig.~\ref{fig:distributed-solvers} are sufficient for solving a broad class of optimization problems  over delay or disruption tolerant networks and further do not rely critically upon the availability or synchronization of any particular compute resources.

	\subsection{Non-distributed implementations} \label{subsec:local}
		The toolset in O-SPC also provides support for four local or non-distributed solver types which organize and implement the associated signal processing system using a single JavaScript enabled web browser as the compute engine. We define an asynchronous implementation protocol in this setting as one for which the behavior of the system state is that of coordinate-wise discrete-time sample-and-hold elements triggered by discrete-time Bernoulli processes. 

		More formally, let $\{\mathcal{I}_{n}\}_{n=1}^{\infty}$ denote a sequence of randomly generated subsets of $\{1,\dots,K\}$ such that for every value of $n$ each $i\in\{1,\dots,K\}$ is included in $\mathcal{I}_{n}$ with probability $p$ and not included with probability $1-p$ independently and independent of $n$. Further, denote $\mathcal{I}^{c}_{n}$ as the set compliment of $\mathcal{I}_{n}$, i.e. $\mathcal{I}^c_n = \{i\in\{1,\dots,K\} \colon i \not \in \mathcal{I}_{n}\}$, and let $I_{\mathcal{I}_n}$ denote the diagonal matrix with ones on the diagonal entries indicated by the index set $\mathcal{I}_n$ and zeros elsewhere.  Then, the update procedure for the state sequence $\dn$ given by
		\begin{eqnarray}
			\d^n = I_{\mathcal{I}_{n}}\left(Gm\left(\d^{n-1}\right)+\f\right) + I_{\mathcal{I}_n^c}\d^{n-1}, & & n \in \N, \label{eq:local-iter}
		\end{eqnarray}
		corresponds to the iterative solver for the signal processing system depicted in Fig.~\ref{fig:signal-flow}(d). Reorganizing the computation and modifying the initial conditions such that the first difference of the signals rather than the signals themselves are being processed results in the incremental solver processing procedure where $\c^{n} = m(\d^{n-1})$ and 
		\begin{eqnarray}
			\d^n = \d^{n-1} + GI_{\mathcal{I}_n}\left(m\left(\d^{n-1}\right) - \c^{n-1} \right), & & n \in \N.\label{eq:local-incr}
		\end{eqnarray} 

		The system initialization and processing procedure for the local solvers discussed hereto and their filtered counterparts are summarized in Fig.~\ref{fig:local-solvers}. In addition, screen captures from O-SPC illustrate the solution obtained to a sparse signal recovery problem wherein the signal processing system was implemented using the incremental filtered solver with $\rho=0.5$ and $p=0.25$. 

		\begin{figure*}[t]
			\setcounter{figure}{2}
			\centering
		  	\centerline{\includegraphics[width=7.1in]{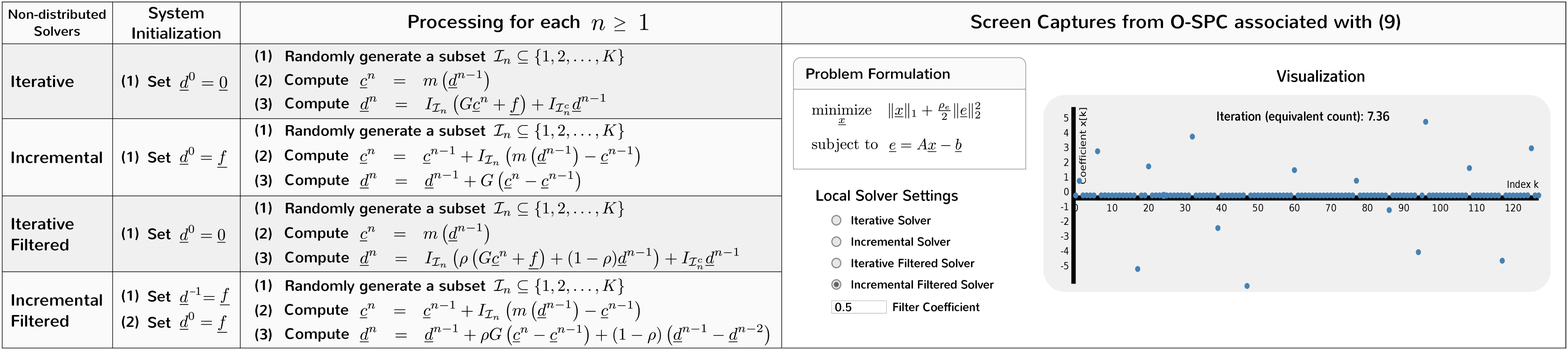}}
		  	\caption{The procedures for system initialization and processing associated with four non-distributed solvers. Screen captures from O-SPC illustrate the solution to a LASSO or basis pursuit denoising problem \cite{cs} obtained by running an incremental filtered solver with filter parameter $\rho = 0.5$ and $p=0.25$.}
			\label{fig:local-solvers}
		
		\end{figure*}

\section{Numerical examples} \label{sec:examples}
	Mathematical optimization typically manifests itself in signal processing applications as either a design tool for optimal parameter selection or a processing stage in the signal chain itself. In this section we provide context and commentary for examples of these types depicted in Figs.~\ref{fig:distributed-solvers} and \ref{fig:local-solvers}. In addition, we present a third and final example related to error correction in transform coding theory solved using O-SPC. The obtained solutions agree with those generated by CVX \cite{cvx}. We conclude with a discussion of the relationships between the specific signal processing systems associated with the~examples.

	\subsection{Sparse signal recovery}
		A well-established approach to recovering a sparse signal measured through an underdetermined linear system that has potentially been corrupted by noise is to solve the LASSO or basis pursuit denoising problem. In particular, this recovery formulation is posed as a regularized least squares problem of the form 
		\begin{eqnarray} \label{eq:ssr}
			\displaystyle \minimize_{\x} 	& \frac{\rho}{2}\|A\x - \y\|_2^2 + \|\x\|_1
		\end{eqnarray}
		where $A\in\R^{m\times n}$ is the linear measurement system, $\y\in\R^m$ is a vector of measurements, $\rho > 0$ scales the objective function, and $\x\in\R^{n}$ is the desired sparse vector. We draw $A$ at random from a Gaussian ensemble to ensure it satisfies the restricted isometry property with high probability \cite{cs}. The solution depicted in Fig.~\ref{fig:local-solvers}, solved using O-SPC, corresponds to $(m,n)=(60,128)$. Problems sizes of the order $(m,n) = (2400,5120)$ were additionally solved, i.e.~ where $A$ has $\approx12$ million non-zero entries.

	\subsection{Non-negative least squares}
		The non-negative least squares problem, which is commonly used as a subroutine in solving more general non-negative tensor factorization problems, is formulated as  the quadratic program
		\begin{eqnarray} \label{eq:nnls}
			\displaystyle \minimize_{\x } 	& 	\frac{1}{2}\|A\x - \y\|_2^2  & \text{s.t. } \x \geq \underline{0}
		\end{eqnarray}
		where $A\in\R^{m\times n}$ is a general linear system, $\y\in\R^m$ is a vector of observations, and $\x\in\R^{n}$ is the desired non-negative vector. For the example depicted in Fig.~\ref{fig:distributed-solvers}, $m$ and $n$ were respectively selected to be $128$ and $60$. Constrained least squares problems such as \eqref{eq:nnls} have immediate application to system design in a number of ways.  For example, in the context of filter design, \eqref{eq:nnls} facilitates the design of filters including peak-constrained least squares filters \cite{lsfilter} with additional non-negativity constraints on the filter taps enabling their use on implementation technologies with unsigned number systems. 
		
	\subsection{Error correction decoding} 
\begin{figure}[t]
			\centering
			\centerline{\includegraphics[width=4.5in]{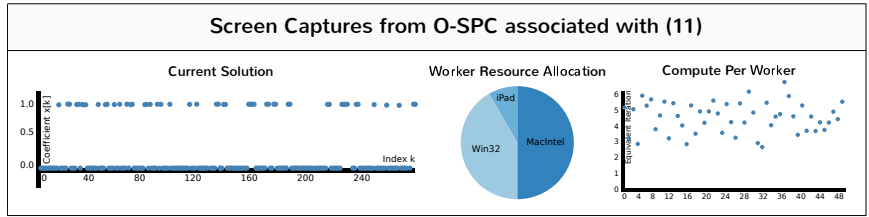}}
			\caption{An illustration of the computed solution of \eqref{eq:ecd} obtained using $50$ distributed workers implementing an iterative filtered solver with parameter $\rho = 0.75$. A breakdown of the workers computational platform distribution and individual contributions to the overall iteration count is also depicted.}
			\label{fig:decoder-example} 
		\end{figure}

		Let $A\in\R^{m\times n}$ denote a linear codebook, i.e. with each column of $A$ denoting a codeword, and consider the recovery of a plaintext vector $\x \in \R^{n}$ from a cyphertext vector $\y \in \R^{m}$ which has been additively corrupted by a $p$-sparse noise vector $\z \in \R^{m}$ according to $\y = A\x + \z$. We cast the recovery procedure as the problem 
		\begin{eqnarray} \label{eq:ecd}
			\displaystyle \minimize_{\x} & \left\|A\x - \y \right\|_{1},
		\end{eqnarray}
		hence decoding a given cyphertext vector in this way corresponds to solving a linear program since \eqref{eq:ecd} may be recast as the standard basis pursuit problem. Furthermore, \eqref{eq:ecd} is guaranteed to identify the correct plaintext vector $\x^\star$ so long as $A$ and the triple $(n,m,p)$ satisfy the conditions provided in \cite{tao}. Figure~\ref{fig:decoder-example} depicts the solution obtained from the numerical experiment outlined in \cite{tao} using the distributed iterative filtered solver presented in this paper where we specifically select the transmitted plaintext vector to be binary valued and round the decoded plaintext vector for further noise suppression. The solver was implemented using $50$ distributed workers. Note that the plaintext vector obtained using \eqref{eq:ecd} is indeed the synthetic plaintext vector before transmission.

	\subsection{Comments on the example signal processing systems}
		The optimization problems \eqref{eq:ssr}-\eqref{eq:ecd} were specifically chosen to underscore the flexibility and generality of the framework in \cite{BLOpt} with respect to the implementation paradigm discussed in this paper. In particular, for the same linear system $A$ and observation vector $\y$, the signal processing system associated with these three problems differ only in the analytic form of the nonlinearity $m(\cdot)$ used in defining the transformed stationarity conditions \eqref{eq:c-and-d}. The coordinatewise nonlinearity $m_{\eqref{eq:ssr}}: \R \to \R$ associated with the sparse signal recovery problem \eqref{eq:ssr} is given by 
		\begin{eqnarray}
			m_{\eqref{eq:ssr}}(x) = \left\{\begin{array}{lr}-x,& |x| \leq 1 \\ x - 2\text{ sign}(x), & |x| > 1\end{array}\right.,
		\end{eqnarray}
		whereas the coordinatewise nonlinearities $m_{\eqref{eq:nnls}}: \R \to \R$ and $m_{\eqref{eq:ecd}}:\R \to \R$ respectively associated with the non-negative least squares problem \eqref{eq:nnls}  and the error correction decoding problem \eqref{eq:ecd} are given by $m_{\eqref{eq:nnls}}(x) = |x|$ and $m_{\eqref{eq:ecd}}(x) = m_{\eqref{eq:ssr}}(-x)$. Each of these nonlinearities (as scalar operators or stacked into an operator from $\Rk$ into itself) satisfy the sufficient condition for convergence in \eqref{eq:passive-everywhere} and thus, for example, the filtered solvers discussed in Subsections~\ref{subsec:distributed} and \ref{subsec:local} may be directly utilized to solve the corresponding problems. We conclude with a remark on the similarity of the complexity associated with solving \eqref{eq:ssr} and \eqref{eq:ecd} in the sense of identifying fixed-points of the algebraic system \eqref{eq:c-and-d} due in part to the relationship between $m_{\eqref{eq:ssr}}$ and $m_{\eqref{eq:ecd}}$. This similarity may not be readily apparent from the optimization problem statements since \eqref{eq:ecd} is a linear program while \eqref{eq:ssr} is convex quadratic, but can be leveraged to efficiently solve both problem instances without replicating the entire problem in the database.

\newpage
\FloatBarrier
\bibliographystyle{IEEEbib}
\bibliography{refs}

\end{document}

%% file: definitions.tex
\def \a  { \underline{a} }
\def \c  { \underline{c} }
\def \d  { \underline{d} }
\def \e  { \underline{e} }
\def \f  { \underline{f} }
\def \g  { \underline{g} }
\def \v  { \underline{v} }
\def \u  { \underline{u} }

\def \x  { \underline{x} }
\def \y  { \underline{y} }
\def \z  { \underline{z} }
\def \R  { \mathbb{R}    }
\def \N  { \mathbb{N}    }
\def \Rk { \mathbb{R}^K  }
\def \Rn { \mathbb{R}^N  }

\def \cn { \{\underline{c}^n\}_{n=0}^{\infty} }
\def \dn { \{\underline{d}^n\}_{n=0}^{\infty} }

\DeclareMathOperator*{\minimize}{minimize}